\documentclass[aps,twocolumn,groupedaddress]{revtex4}

\newif\ifpdf
\ifx\pdfoutput\undefined
\pdffalse 
\else
\pdfoutput=1 
\pdftrue
\fi

\ifpdf
\usepackage[pdftex]{graphicx}
\else
\usepackage{graphicx}
\fi
\usepackage{hyperref}

\begin{document}

\ifpdf
\DeclareGraphicsExtensions{.pdf, .jpg}
\else
\DeclareGraphicsExtensions{.eps, .jpg}
\fi

\def\hslash{\hbar}
\def\imag{i}
\def\grad{\vec{\nabla}}
\def\div{\vec{\nabla}\cdot}
\def\curl{\vec{\nabla}\times}
\def\DDt{\frac{d}{dt}}
\def\ddt{\frac{\partial}{\partial t}}
\def\ddx{\frac{\partial}{\partial x}}
\def\ddy{\frac{\partial}{\partial y}}
\def\lap{\nabla^{2}}
\def\divv{\vec{\nabla}\cdot\vec{v}}
\def\gradS{\vec{\nabla}S}
\def\vvec{\vec{v}}
\def\wc{\omega_{c}}
\def\<{\langle}
\def\>{\rangle}
\def\Tr{{\rm Tr}}
\def\Csch{{\rm csch}}
\def\Coth{{\rm coth}}
\def\Tanh{{\rm tanh}}
\def\g2{g^{(2)}}


\title{Quantum initial value representations using approximate Bohmian trajectories}


\author{Eric R. Bittner}
\affiliation{Department of Chemistry and Center for Materials Chemistry \\
University  of Houston, Houston, TX 77204}


\date{\today}

\begin{abstract}
Quantum trajectories, originating from the de Broglie-Bohm (dBB) hydrodynamic
description of quantum mechanics, are used to construct time-correlation 
functions in an initial value representation (IVR).  The formulation is fully
quantum mechanical and the resulting equations for the 
correlation functions are similar in form to their semi-classical analogs but do 
not require the computation of the stability or monodromy matrix or 
conjugate points.   We then move to a {\em local} trajectory description 
by evolving the cumulants of the wave function along each individual path.
The resulting equations of motion are an infinite hierarchy, which we truncate 
at a given order.   We show that time-correlation functions
computed using these approximate quantum trajectories can be used to 
accurately compute the eigenvalue spectrum for various potential systems.
\end{abstract}


\maketitle

\section{Introduction}

Over the past few years there has been significant effort in the
development of the so-called hydrodynamic or quantum trajectory based
approach to quantum mechanics.
\cite{rabitz98a,qtm1,mayor,bw1,wb1,erb,maddox,maddox3,irene1,irene2,irene3,irene4,hw,tw1,bwreview,maddox2,makri}
This formulation is based upon Bohm's ``hidden variable''
 representation~\cite{bohm,debroglie,holland}
 is initiated by substituting the amplitude-phase decomposition of the time-dependent
quantum wavefunction,
$\psi(y,t) = R(y,t)\exp(i S(y,t)/\hbar)$, into the time-dependent  Schr{\"o}dinger equation,
\begin{eqnarray}
i\hbar\frac{\partial\psi}{\partial t} = \left(-\frac{\hbar^2}{2m}\nabla^2 + V\right)\psi
\end{eqnarray}
Substituting Eq. 1 into Eq. 2 and separating the  real and imaginary components
yields equations of motion for $\rho$ and $S$,
\begin{eqnarray}
\frac{\partial \rho}{\partial t} &=& -\nabla \cdot(\rho \nabla S)/m \label{cont1} \\
\frac{\partial S}{\partial t} &=& -\frac{\nabla S\cdot\nabla S}{2m} + V  
-\frac{\hbar^2}{2m}\frac{1}{\sqrt{\rho}}\nabla^2\sqrt{\rho} = 0\label{qhj1}
\end{eqnarray}
where $\rho = |R|^2$ is the probability density and $S$ is the phase.  These 
last two equations can be easily identified as the continuity equation (Eq.~\ref{cont1})
 and the quantum 
Hamilton-Jacobi equation (Eq.~\ref{qhj1}).  In Eq.~\ref{qhj1} we can 
identify three contributions to the action. The first two are the 
kinetic and potential energies and the third is the 
quantum potential $Q$ \cite{bohm,holland}.
This term is best described as a {\em shape kinetic energy} since
it is determined by the local curvature of the quantum density. 
In numerical applications, computation of the quantum potential is frequently rendered more accurate if derivatives are evaluated using the log-amplitude
$C = \log(R)$
 ($C$ is also referred to as the $C$-amplitude).  In terms of derivatives of this amplitude, the quantum potential is
$Q = -\hbar^2 (\nabla^2 C + (\nabla C)^2)/2m$.

Eq. ~\ref{cont1} and \ref{qhj1} are written in terms of partial derivatives 
in time.  In order to compute the time-evolution of $\rho$ (or equivalently $C$)
 and $S$ along a give path, $x(t)$ we need to move to a 
 moving Lagrangian frame by 
 transforming $d_t = \partial_t + \dot x(t)\partial_x$. 
The Lagrangian form of the hydrodynamic equations of motion resulting from this analysis are given by:
\begin{eqnarray}
\frac{d\rho}{dt} &=& -\rho\nabla\cdot v \\ \label{cont}
\frac{dv}{dt}&=& -\frac{1}{m}\nabla(V+Q)\\ \label{newton}
\frac{dS}{dt} &=& \frac{1}{2m}(\nabla S)^2 - (V+Q) \label{qhj}
\end{eqnarray}
in which the derivative on the left side is appropriate for calculating the rate of change in a function along a fluid trajectory. 
 Eq.~\ref{newton} is a Newtonian-type equation in which the flow acceleration is produced by the sum of the classical force, $f_c = -\nabla V$,
 and the quantum force is $f_q=-\nabla Q$

The quantum trajectories obey two important noncrossing rules: (1) they cannot cross nodal surfaces (along which $\psi=0$ ); (2) they cannot cross each other.  (In practice, because of numerical inaccuracies, these conditions may be violated.)
Because of these two conditions,
the quantum trajectories are very different from classical paths and represent
a geometric optical rendering of the evolution of the quantum wave function.

The primary difficulty in implementing a quantum trajectory based approach
is in constructing the quantum potential.    In spite of 3-4 years of intense effort by
a number of groups\cite{rabitz98a,qtm1,mayor,bw1,wb1,erb,maddox,hw,tw1,bwreview,maddox2}  
, including our own, there has yet to emerge a satisfactory
and robust way to compute $Q$ and the quantum force $-\grad Q$ directly from
the quantum trajectories in multidimensional systems.
This severely limits the applicability of a quantum
trajectory based approach for computing exact quantum dynamics for all but the most trivial
systems.   

In this paper, we show that quantum trajectories can be used to 
construct exact quantum initial value representations (IVR) of correlation functions.
The basic equations are extremely straightforward to derive and follow directly 
from a Bohm trajectory based description.  The heart and sole of our approach, 
however, is the use of an cummulant/derivative propagation scheme~\cite{tw2} that 
generates {\em local} equations of motion for a given quantum trajectory and 
all the derivatives of $C$ and $S$ along the path, allowing us to compute 
the correlation function on a trajectory by trajectory basis.   
In our numerical implementation of this approach, we examine a highly anharmonic
one dimensional system (an inverted gaussian well) and find that the Bohm-IVR
approach is surprisingly accurate in computing correlation functions even though 
at long-times the quantum trajectories themselves show crossings.  

\section{The Bohm Initial Value Representation}
While in principle, the quantum wavefunction or the density matrix tells
us everything we can know about a given physical process, often what we are interested in
is a correlation function or an expectation value.  
This can be equivalently cast as either a
time-correlation or a spectral function via Fourier transform:
$$
\<\hat{A}(\omega)\>  = {\cal F}[\<\hat{A}(t)\hat{A}\>].
$$
Where
\begin{eqnarray}
\<\hat{A}(t)\hat{A}\> &=& \<\psi_o|e^{+iHt/\hbar}\hat{A}e^{-iHt/\hbar}\hat{A}|\psi_o\>
\label{cor}
\end{eqnarray}
is the time-correlation function for some quantum operator $\hat{A}$ following
initial state $\psi_o$ evolving under Hamiltonian, $H$.  

\begin{figure}[h]
\includegraphics{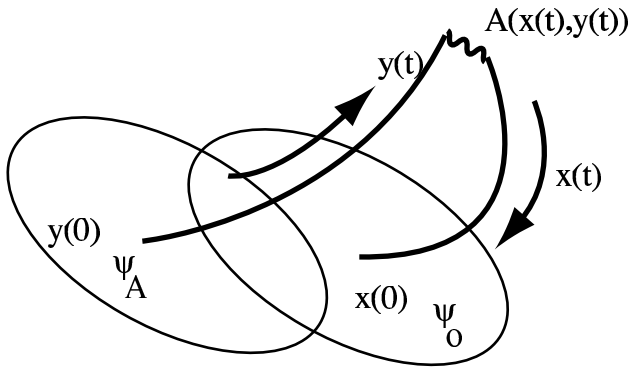}
\caption{Schematic representation of how correlation functions are 
constructed within the Bohm initial value representation of Eq.\ref{bohmIVR}.
We start off with an initial wavefunction, $\psi_A=\hat{A}\psi$,
 sample an ensemble of starting points $\{y(0)\}$
and evolve  quantum trajectories out to time $t$.
 Simultaneously, one evolves a series of quantum trajectories \emph{backwards} 
in time starting with $\psi_o^*$ as the initial wave function.  
The two trajectories are connected at time $t$ via the matrix elements 
of $\hat A$ at the end points, $A(x(t),y(t))$.   
 }\label{prop}
\end{figure}

Within the Bohmian framework, the synthesis of a quantum wavefunction along a quantum trajectory, $x(t)$, is
given by
\begin{eqnarray}
\psi(x,t) &=& \int dx(t) \delta(x-x(t))
J^{-1/2}(x(t),x(0)) \nonumber \\
&\times&
e^{ iS(x(0),x(t))/\hbar}\psi_o(x(0)).
\end{eqnarray}
Here, 
\begin{eqnarray}
J(x(t),x(0))=e^{+\int_0^t \nabla\cdot v ds}
\end{eqnarray}
is the Jacobian for the transformation of a
volume element $dx(0)$ to $dx(t)$ along path $x(t)$,  $dx(t)=J(x(0),x(t)) dx(0)$.
Its time evolution follows directly from the continuity  equation with initial condition
$J(x(0),x(0)) = 1$.   Likewise, $S$ is the action integral integrated over the
quantum trajectory connecting $x(0)$ and $x(t)$. 
Using this, we can rewrite Eq.~\ref{cor} as an integral over
starting points for Bohmian trajectories:
\begin{widetext}
\begin{eqnarray}
\<\hat{A}(t)\hat{A}\> =
\int dx(0) \int dy(0) A(x(t),y(t))J_x^{1/2}J_y^{1/2}e^{ -iS(x(0),x(t))/\hbar}\psi^*(x(0)) e^{ iS_A(y(0),y(t))/\hbar}\psi_A(y(0))
\label{bohmIVR}
\end{eqnarray}
\end{widetext}
where $|\psi_A\>=\hat{A}|\psi\>$ and we assume that $\hat{A}$ is a scalar 
quantum operator
with matrix elements $\<x|\hat{A}|y\> = A(x,y)$.
In this expression, the forward-going 
quantum path along $y(0)$ to $y(t)$ is subject to a quantum potential
derived from the evolution of $\psi_A(t)$ 
(with action integral $S_A(y(0),y(t))$
and Jacobian $J_y$), where as the reverse path along $x(0)$ to $x(t)$ has a
quantum potential derived from the evolution of $\psi(t)$ (with action integral $S(x(0),x(t))$
and Jacobian $J_x$.)
Thus, even if $x(0)=y(0)$
the forward path ($y(s)$) will be very different than the reverse path.   We can interpret this
in the following way.  At time $t=0$, we act on the initial wave function with operator
$\hat A$.  If $\psi$ is not an eigenstate of $\hat A$, then we have a new wave function
$\hat A|\psi\> = |\psi_A\>$ which evolves out to time $t$ where we act again with $\hat A$
and $y(t)$ is instantly scattered to $x(t)$ where it evolves back in time under the
reversed evolution of $\psi(t)$.  The IVR correlation function then measures the
probability amplitude that the final forward-backward state at $x(0)$ was a member of the initial state at $y(0)$.
  This is illustrated schematically in Fig.~\ref{prop}.

So far, {\em no approximations have been made}.    The simplest 
correlation function is that the autocorrelation function of the wavefunction 
which monitors the overlap between the initial wavefunction $\psi_o$ and the
time-evolved wavefunction.
\begin{eqnarray}
c(t) &=& \<\psi(0)|\psi(t)\> \nonumber \\
&=& \int dx(0) \psi_o^*(x(t)\psi_o(x(0))J^{1/2} e^{iS/\hbar}\label{exactivr}
\end{eqnarray}
Here, $\psi_o(x)$ is the initial wave function. 
This is directly 
analogous to the semi-classical IVR developed extensively
over the past number of years.
\cite{makri,gr1,mchild,makri02,miller03}
In the SC version, one estimates the quantum propagator via the 
well known Van Vleck expression~\cite{vvleck}.
 \begin{eqnarray}
c(t) 
  &\approx&\int dx(0) \psi^*_o(x(t)) K_{SC}(x(t),x(0);t) \psi_o(x(0)).\label{scivr}
\end{eqnarray}
Here, $K_{SC}(x(t),x(0);t)$ is the semi-classical propagator
\begin{eqnarray}
K_{SC} =  \sum_{cl.\, paths} 
\frac{{\| C\|^{1/2}}}{(2 \pi i \hbar)^{D/2}} e^{iS_{cl}-i\kappa \pi/2} 
\end{eqnarray}
where the summation is over all 
classical paths connecting $x(0)$ to $x(t)$ 
with classical action $S_{cl}$, 
\begin{eqnarray}
S_{cl} = \int_0^t \frac{m}{2}\dot{x}^2(s) - V(x(s)) ds,
\end{eqnarray}
$C$ is the matrix of the negative second variations of the classical action with the end points,
\begin{eqnarray}
C&=& -\frac{\partial^2 S_{cl}}{\partial x(0) \partial x(t)} 
\end{eqnarray}
and $\kappa$ gives the Morse index obtained by counting the number of conjugate points along each 
trajectory.  
What is appealing in the Bohm-IVR expression above is that 
one does not need to compute $C$ or $\kappa$ along
each trajectory.   This is  not surprising since we are dealing 
with a purely quantum mechanical object and $C$ and $\kappa$ are 
semi-classical quantities.  

A closer connection between Eq.~\ref{scivr} and Eq.~\ref{exactivr} can be drawn by 
writing $C$ as 
\begin{eqnarray}
C&=& -\frac{\partial p(t)}{\partial x(0)} 
 = -J \frac{\partial^2 S_{cl}}{\partial x(t)^2}
\end{eqnarray}
Thus, the SC-IVR expression can be written as
\begin{widetext}
\begin{eqnarray}
c_{sc}(t)  &=&
\frac{1}{(2 \pi i \hbar)^{D/2}}
\int dx(0) 
\sum_{cl.\, paths} 
J^{1/2}{\left|\frac{\partial^2 S_{cl}(t)}{\partial x^2(t)}\right|}^{1/2}
e^{iS_{cl}/\hbar -i\kappa \pi/2}
\psi^*_o(x(t)) \psi_o(x(0)).\label{scivr2}
\end{eqnarray}
\end{widetext}
Comparing Eq.~\ref{scivr2} to Eq.~\ref{exactivr} we first note the appearance of 
$J^{1/2}$  in both expressions.  
Here, as in the case above,  $J$ is the Jacobian for the transformation of a 
volume element $dx(0)$ to $dx(t)$ along the \emph{classical path} $x_{cl}(t)$.
What is most striking, however, is that in the 
exact Bohm IVR expression one has a \emph{single trajectory connecting 
each initial point to each final point} where as in the semi-classical case, we 
have to consider every possible trajectory that starts at $x(0)$.  In the semi-classical 
case, this can take the form of a phase-space integration, as in the widely used
Herman-Kluk expression\cite{hk,hkd,mh86,gr1,mchild,batista,makri02,gara-light}. 
%

As mentioned above, one of the primary difficulties encountered in developing 
numerical methods based upon the Bohm trajectory approach is in computing 
the quantum potential, $Q$, and its derivatives on a generally unstructured 
mesh of points evolving according to the Bohmian equations of motion.  Moreover, the
trajectories themselves can be highly erratic and kinky requiring that one 
use very accurate numerical integration methods to achieve an accurate 
solution  of the equations of motion.~\cite{makri}  Consequently, obtaining an accurate
estimate of the time-evolving wavefunction (via the synthesis equation above) 
is quite difficult.   On the other hand, one may be able to get by with a less robust and 
less exact calculation of the quantum trajectories if one is only interested in computing
time-correlation functions.  
We next turn our attention towards a  hierarchy
approach for generating approximate quantum trajectories.


\section{Derivative Propagation}
Recently, Trahan, Hughes, and Wyatt~\cite{tw2} introduced an interesting approach for computing
quantum trajectories by simultaneously evolving the spatial derivatives of both $C$ and
$S$ along a given trajectory.   This notion is similar in spirit to the hydrodynamic moment
expansions of the density matrix\cite{maddox,maddox3,irene1,irene2,irene3,irene4} 
and follows directly by
taking derivatives of $\partial C/\partial t$ and $\partial S/\partial t$ with respect to $x$
and transforming to the moving Lagrangian representation.   Following this notion, the
time-evolution of the $n$-th spatial derivatives of $C$ and $S$ at $x_k(t)$ are given by
\begin{widetext}
\begin{eqnarray}
\frac{dC_k^{(n)}}{dt}&=&-\frac{1}{2m}\left(S_k^{(n+2)} + 2\sum_{j=0}^n
\left(\begin{array}{cc} n \\ j \end{array}\right)
S_k^{(n+1-j)} C_k^{(j+1)}\right) + S_k^{(1)}C_k^{(n+1)}/m\\
\frac{dS_k^{(n)}}{dt}&=&-\frac{1}{2m}\left(\sum_{j=0}^n
\left(\begin{array}{cc} n \\ j \end{array}\right)
S_k^{(j+1)}S_k^{(n+1-j)}\right) \nonumber \\
&+&\frac{\hbar^2}{2m}\left(C_k^{(n+2)}
+\sum_{j=0}^n
\left(\begin{array}{cc} n \\ j \end{array}\right)
C_k^{(j+1)}C_k^{(n+1-j)}\right)
-V_k^{(n)} + S_k^{(1)}S_k^{(n+1)}/m
\end{eqnarray}
\end{widetext}
Our notation is that $f_k^{(n)} = \nabla^n f$ evaluated at $x_k(t)$ and 
$(^n_j)$ is the binomial coefficient.
These equations, along with $dx_j/dt = S_j^{(1)}/m$ are 
an infinite hierarchy of equations for both $C$ and $S$ and their derivatives
evaluated along the path $x_j(t)$.  
 In this hierarchy, the  equation for the $n^{th}$ order derivative
of $C$ involves up to the $n+2$ and order derivative of $S$ and up to the $n+1$ 
order derivative of $C$.  Likewise, the equation of motion for the 
$n^{th}$ order derivative of $S$ involves up to 2 higher order terms of $C$ and 
one higher order term of $S$.   The $C_k^{(n)}$ and $S_k^{(n)}$ terms 
are in essence $n^{th}$-order cumulants of the wavefunction evaluated at $x_k(t)$. 
Note also that the hierarchy equations are {\emph local} in that they do not
couple different trajectories.  Each trajectory can be computed as an independent 
quantity and the task of computing multiple quantum trajectories can be
easily distributed over multiple machines.  

Imbedded in these equations
of motion is the mechanism by which complex structure emerges 
within the wave function since even simple wave functions with few cumulants 
will rapidly evolve into wave functions with high order cumulants. 
Thus, in order to close the equations of motion, we need to introduce an
artificial boundary condition  which limits the extent of the hierarchy.  The 
most straight-forward scheme is to simply require any term, $C_j^{(n)}$ or
$S_j^{(n)}$ with $n > n_{order}$ to be set to zero in the initial equations of motion. 
One can also introduce ``absorbing boundary conditions'' by damping 
higher-order terms, e.g. $dS_j^{(n)}/dt = -\lambda S_j^{(n)}$.   

To compute correlation functions, we first start off by sampling
$\psi_o$ on a uniform grid of DVR points such that $\psi_{on} = \psi_o(x_n(0)) w^{1/2}_n$
where $w^{1/2}_n$ is the gaussian quadrature weight associated with the $n$-th
grid point.
Thus, the correlation function can be evaluated by gaussian quadrature over
 a grid defined by the initial positions of the Bohm trajectories.  For example, for the
auto-correlation function of the wavefunction:
\begin{eqnarray}
c(t) = \sum_n w_n\psi_{o}^*(x_n(t))\psi_{o}(x_n(0))J^{1/2}_n e^{iS_n/\hbar}
\label{cdvr}
\end{eqnarray}
where $J_n$ and $S_n$ are the Jacobian and quantum action along $x_n(s)$ which starts
at $x_n(0)$ and ends at $x_n(t)$.

Written in terms of the derivative terms above, the 
wave function correlation becomes simply
\begin{eqnarray}
c(t)  = \sum_{j=1}^{N} w_j e^{-2 C_j^{(0)}(t)} e^{iS_j^{(0)}(t)/\hbar}\psi_o^*(x_j(t))\psi_o(x_j(0))
\end{eqnarray}
where $C_j^{(n)}$ and  $S_j^{(n)}$are computed using the above hierarchy equations 
along a finite set of trajectories.  As each trajectory is independent, the sum can 
evaluated by distributing the trajectory calculations amongst several nodes, each running
a single quantum trajectory.  
This last expression  along with its implementation 
comprise the central results of this paper.  In the following 
section we examine a series of one dimensional test cases.

\section{Sample Calculations}

As discussed above, quantum interference in the wave function leads to 
very complex structure in the quantum potential. Even in the 
simple double-slit experiment example, the quantum potential 
due to the interference between the two gaussian components 
of the wave function varies rapidly as the underlying quantum 
trajectories avoid one another. ~\cite{holland}  Consequently, 
in numerical applications of the quantum trajectory approach, 
nodes and interferences in the wave function have present considerable
hurdles in the development of generalized multi-dimensional 
applications of this method.    

\begin{figure}
\includegraphics[width=3.0in]{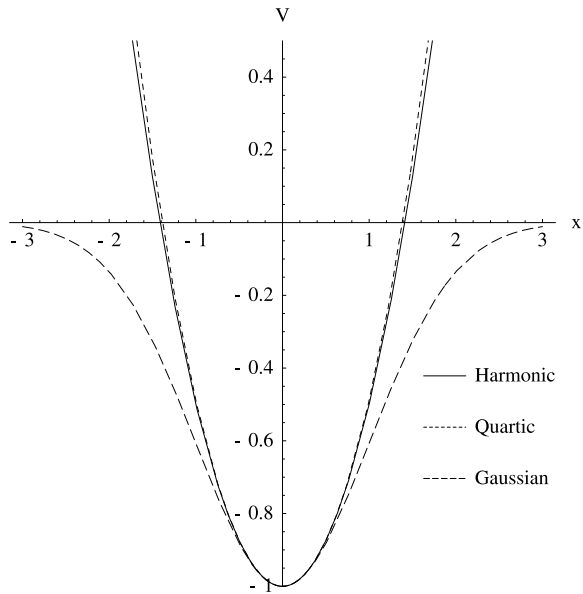}
\caption{Potential functions used in comparative examples.
(Key: $-$ = Harmonic ($V_{HO})$; $\cdot\cdot\cdot =$ Quartic,
($V_{QO}$); $-\cdot-$=Gaussian well ($V_G$).) 
}\label{potens}
\end{figure}

Here, we take three related test cases with increasing anharmonicity and
test the ability of the Bohm-IVR approach with derivative propagation
to obtain accurate correlation functions.   In each case we consider an 
electron ($m=1$) in one-dimensional potentials,  chosen such that the 
curvature at the bottom is identical in each case. 
 \begin{eqnarray}
V_{HO} &=&x^2/2  - 1\\
V_{QO} &=& x^2/2 + 0.01 x^4 -1 \\
V_{G} &=& -e^{-x^2/2}
\end{eqnarray}
$V_{HO}$ is the harmonic well, $V_{QO}$ is harmonic well perturbed 
by a small quartic term, and $V_G$ is an inverted gaussian well.  
In each case, the initial state $\psi(x) = \exp(c+is/\hbar)$
is gaussian with $c$ and $s$ given by 
\begin{eqnarray}
c(x) = \frac{1}{4}\log(\beta/\pi) - \beta(x-x_o)^2 \\
s(x) = 0
\end{eqnarray}
with $x_o = 1$ and $\beta = 1$.  (Note, atomic units used throughout.)
 A plot of each potential along with the location of the lowest  energy 
bound states is show in Fig.~\ref{potens}.

\subsection{Harmonic Oscillator}
\begin{figure}[t]
\includegraphics[width=3.0in]{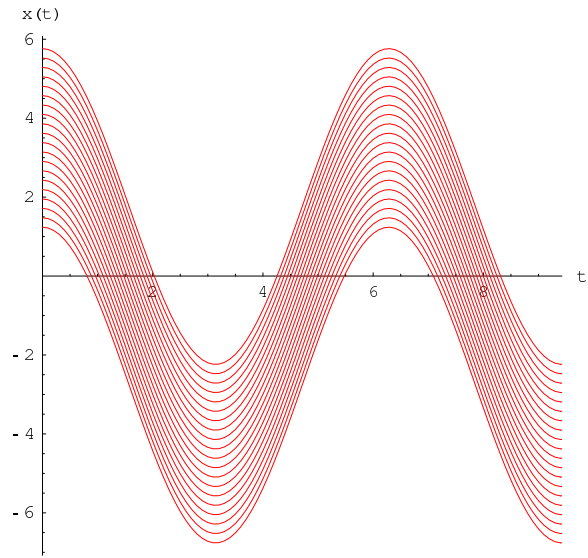}
\caption{Quantum trajectories for harmonic oscillator computed using 
derivative propagation approach.}\label{fig1}
\end{figure}

Harmonic systems are particularly useful as test cases since 
the an initial gaussian wave function retains its gaussian form for all time.  
Furthermore, any quantity of interest can be computed analytically
thereby providing a useful benchmark for the method.  
In Fig.~\ref{fig1} we show the quantum trajectories computed via the 
derivative propagation scheme outlined above. 
The trajectories display the coherent 
oscillation one expects for Bohm trajectories in this system.  This is not 
surprising since a low order 
truncation of the derivative propagation equations is 
perfectly valid for this case. 
 These agree perfectly
with the analytical results for this system and we can integration the derivative
propagation equations out to very long times without significant loss of 
accuracy. The initial points were chosen as evenly spaced 
gaussian quadrature points.

\begin{figure}[t]
\includegraphics[width=3.0in]{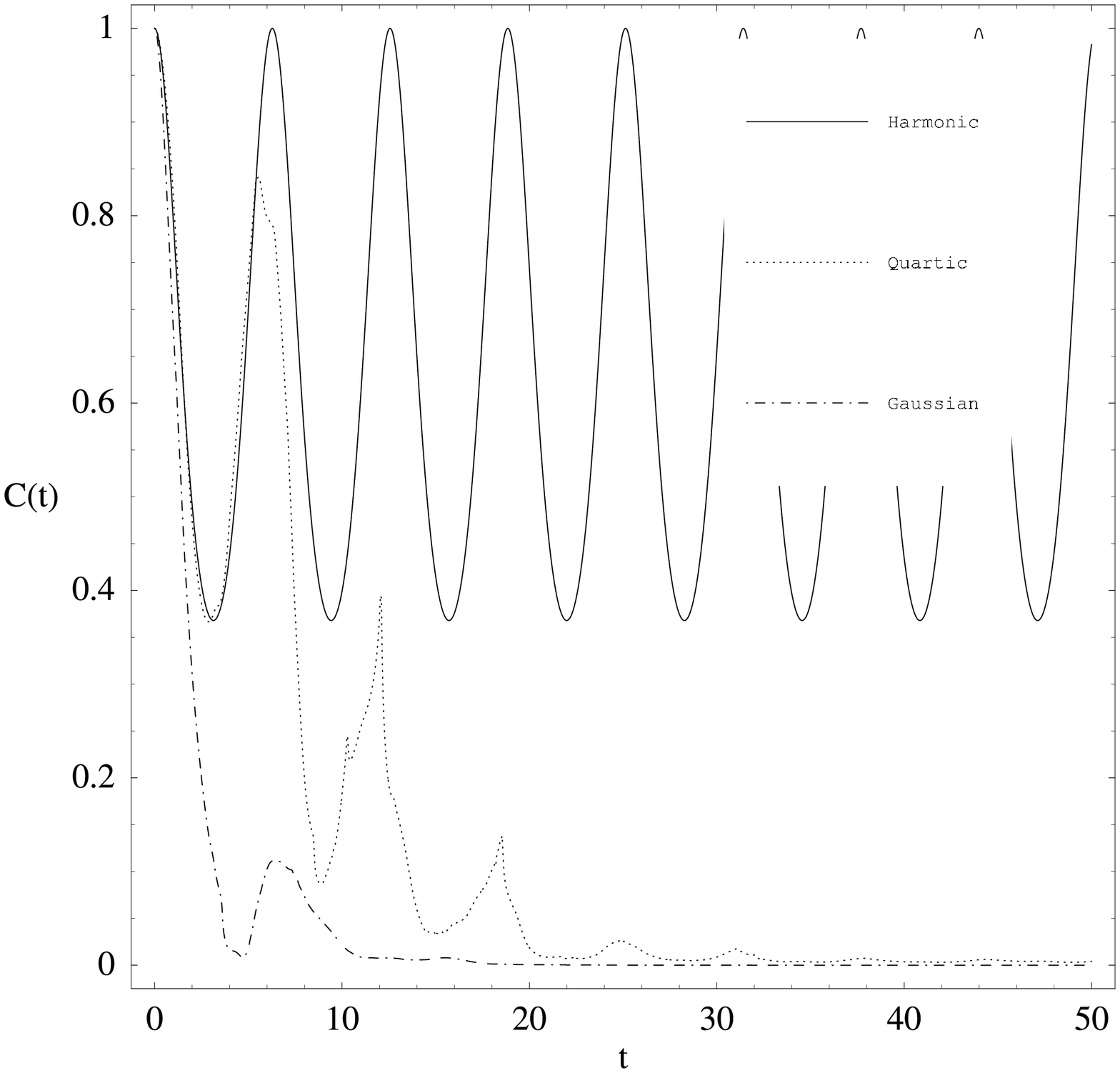}
\includegraphics[width=3.0in]{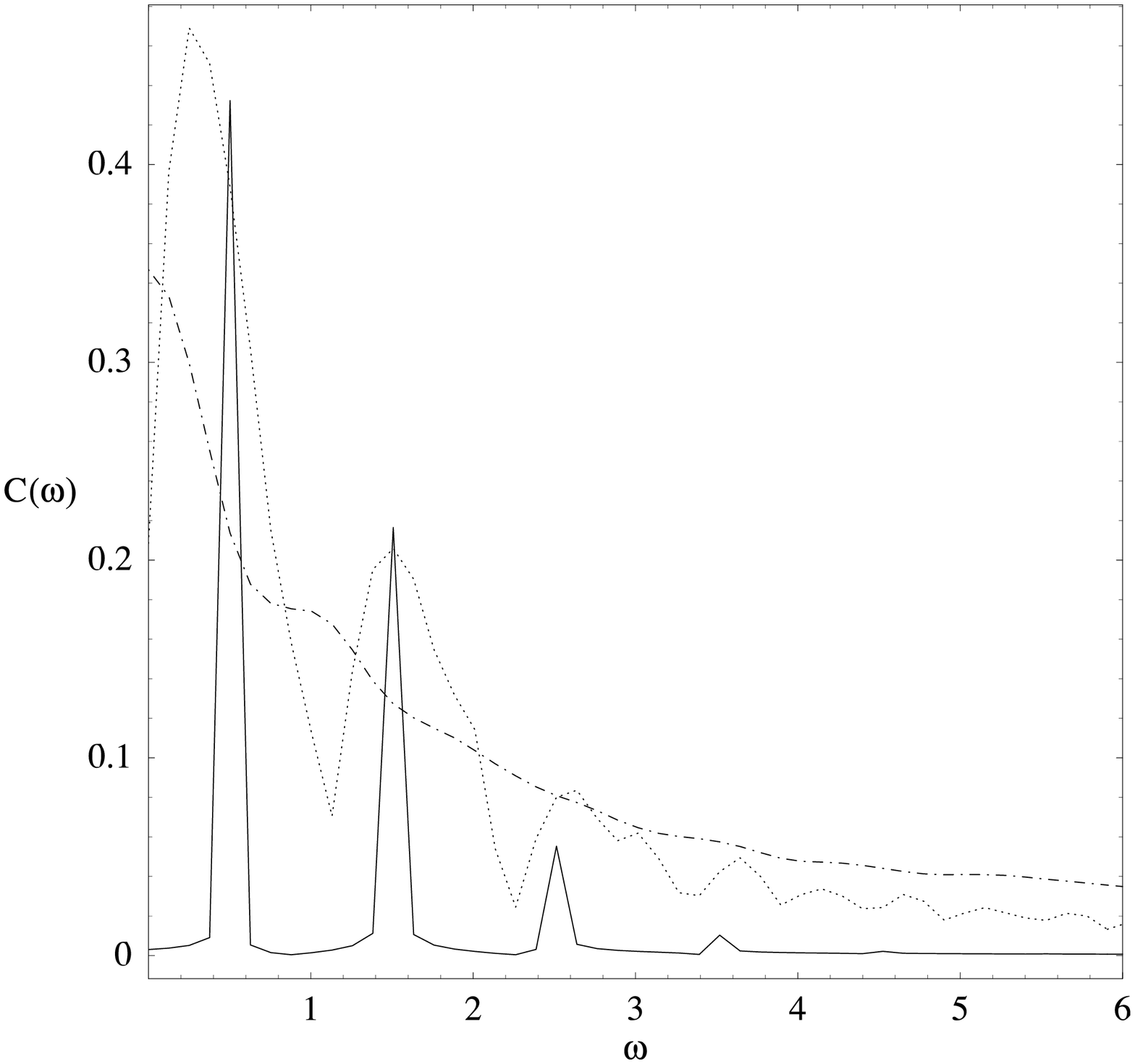}
\caption{Correlation function  and its spectral representation for a 
gaussian wavepacket in various potentials
 computed using Bohm-IVR/derivative propagation approach.  (Key: $-$ = Harmonic ($V_{HO})$; $\cdot\cdot\cdot =$ Quartic,
($V_{QO}$); $-\cdot-$=Gaussian well ($V_G$).)  }\label{ctplot}
\end{figure}

The solid curve in 
Fig.~\ref{ctplot} shows the time correlation function 
of the wavefunction and its Fourier transform.  
Here, we performed the calculation 
out to $t =  20 \pi$ corresponding to 10 complete oscillation cycles.  
Both $C(t)$ and its spectral representation $\tilde{C}(\omega)$ agree perfectly
with analytical results with $\tilde{C}(\omega)$ accurately mapping out the
harmonic progression of the energy levels in this well $\omega_n = \omega(n+1/2)$
as well as the expansion coefficients for the
projection of the initial wavefunction on to the 
eigenstates of the well, $c_n  = |\<n|\psi_o\>|^2$.  For the case at hand,  the 
first 4 coefficients are
$c = \{ 0.606531 , 0.303265 , 0.0758163, 0.0126361 \}$ and agree nicely
with the roughly 8:4:1 ratio of the first three peaks in $C(\omega)$. 

\subsection{Quartic Perturbed Harmonic}

In this case, we apply a slight anharmonic squeeze to the harmonic well. 
In an anharmonic system, an initial gaussian wave function retains its gaussian 
form for short period of time before bifurcating into a multitude of components 
due to interferences.  Such features are most likely to occur where the underlying 
classical trajectory crosses a conjugate or focal point.  Such crossings 
are strictly forbidden for quantum Bohmian trajectories and lead to 
rapidly varying quantum forces over a short spatial region.  Consequently, 
interference effects in anharmonic systems has proven to be somewhat of a bugbear 
for implementing quantum trajectory approaches based upon fitting 
the quantum potential (and its derivatives) to a simple polynomial over a 
set of quantum trajectories.    

The derivative propagation trajectories also retain their coherences
for a short period of time.  However, after only one or two oscillations, crossings
become more frequent and the ensemble of trajectories degrades into an 
incoherent mess.  Had we performed this with the moving-weighted least squares
(MWLS) or other methods which extracts the quantum potential 
from an ensemble of trajectories, the results would have been cataclysmic. 
Our experience has indicated that even small 
errors in constructing the quantum force from the ensemble can lead to 
a rapid amplification of error in the propagation. 
Remarkably, even though the {\em ensemble} has apparently lost coherence
amongst its members, the individual trajectories exhibit non-classical deflections, 
and quasi-periodic behavior one expects from truly quantum trajectories.  

Turning towards the autocorrelation and its Fourier transform.  $C(t)$ 
shows some recurrences at $t \approx 2\pi n$ as in the harmonic case, but these 
rapidly die out after 5-6 beats.  This limits our ability to resolve the 
eigenenergies in the Fourier transform.  Nonetheless, at least 4 peaks 
can be distinguished. 

\subsection{Inverted Gaussian Potential}

\begin{figure}
\includegraphics[width=3.0in]{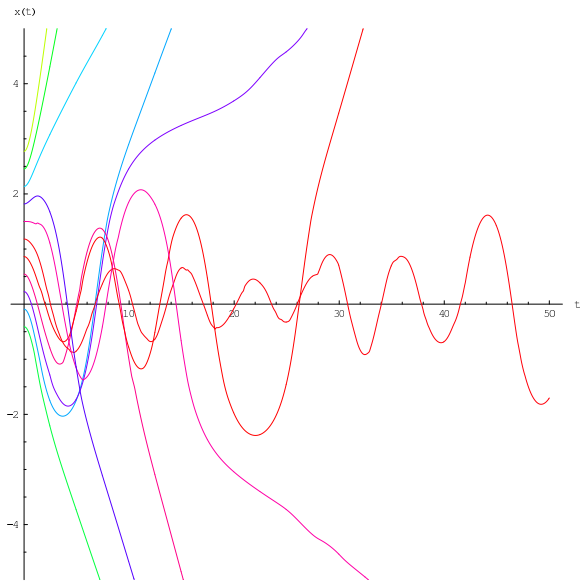}
\caption{Quantum trajectories for inverted gaussian well example.  }\label{gaustraj}
\end{figure}

In this final example, $V_G$ is an inverted gaussian 
and supports only  two eigenstates (at $E = -0.59386$ and $E = -0.0356576$).
In this example we used 30 trajectories 
since the majority of the trajectories near the 
leading edges of the distribution leave the potential region 
almost immediately.   
In Fig.~\ref{gaustraj} we show 10 trajectories that started 
in the center of the initial wavepacket.   As time progresses, many of these
eventually escape the potential well leaving only one of the original 30 trajectories in the well. 

The long-time accuracy of the ensemble is not so important since we are ultimately interested in 
computing $C(t)$.   This we show in Fig.~\ref{ctplot} along with its Fourier transform. 
One can clearly identify 6 recursion peaks spaced roughly every 0.35 ps.  These correspond
 to the nearly harmonic motion of the trajectories closest to the center of the 
ensemble.

\section{Discussion}

For low dimensional systems, there are arsenals of robust and computationally efficient 
methodologies for computing exact quantum dynamics for both time dependent and 
time-independent systems.  However, these methods are severely limited 
by dimensionality since the computational effort and 
storage required to perform an exact quantum calculation 
increases almost exponentially with dimensionality.  
Thus, approximate methods, such as those introduced here, provide a 
viable path way towards higher dimensions with increasing complexity.  
While we have focused entirely upon one-dimensional examples, the Bohm-IVR
and derivative propagation 
equations are exact and can be easily implemented in a multi-dimensional 
setting.

{\em What do the individual trajectories mean? }  Are the trajectories 
computed via the derivative propagation methods {\em really} Bohm trajectories?
The answer, I believe is a definite yes and no.   Yes, they are Bohm trajectories, 
but not for a initial single gaussian wavepacket evolving in a given potential. 
They correspond to trajectories arising from a diverging set of wave functions 
where each wave function component is constrained to have a particular functional 
form dictated by truncating the hierarchy.   In essence, by truncating the 
hierarchy, one introduces dephasing and decoherence into the dynamics 
of a quantum system.   This is evident in the time-correlation function for the 
particle in the harmonic + quartic  potential.  The time correlation
function for a fully coherent wavepacket should show stronger recurrences 
and  eventually be able to resolve the eigenenergies of the system upon 
Fourier transform.   
One evidence for this is that for bound systems other than the  harmonic oscillator,
the autocorrelations we compute using the derivative propagation decay to zero after a few recurrences, and never recover coherence.    

For exact quantum mechanics, this implicit loss of coherence at long-times clearly undesirable. 
However, for condensed phase systems, especially in cases where we have a low dimensional 
quantum subsystem in contact with a high-dimensional ``bath'' of oscillators, reduction of the 
density matrix of the subsystem to a gaussian form in the coherences   occurs  spontaneously
due to decoherence.~\cite{maddox,irene3,irene4}. Consequently, so long as the ``physical'' decoherence
time is smaller than the ``truncation induced '' decoherence time,  
time-correlation functions computed using the 
Bohm-IVR approach we describe herein are expected to quite accurate. 

\begin{acknowledgments}
This work was supported by the National Science Foundation and the 
Robert A. Welch Foundation. The author wishes to acknowledge discussions 
with Jeremy Maddox regarding this work and Prof. Bob Wyatt for providing us
with a preprint of Ref.\cite{tw2}.
\end{acknowledgments}

\end{document}